\documentstyle[prl,aps,epsf]{revtex}
\draft
\begin{document}

\twocolumn

\title{Anderson transitions in three-dimensional disordered systems with
randomly varying magnetic flux}
\author{Tohru Kawarabayashi}
\address{I. Institut f\"{u}r Theoretische Physik, Universit\"{a}t Hamburg,
Jungiusstra{\ss}e 9, D-20355 Hamburg, Germany \\ and \\
Institute for Solid State Physics, University of Tokyo,
Roppongi, Minato-ku, Tokyo 106, Japan}
\author{Bernhard Kramer}
\address{I. Institut f\"{u}r Theoretische Physik, Universit\"{a}t Hamburg,
Jungiusstra{\ss}e 9, D-20355 Hamburg, Germany}
\author{Tomi Ohtsuki}
\address{Department of Physics, Sophia University,
Kioi-cho 7-1, Chiyoda-ku, Tokyo 102, Japan}

\date{\today}
\maketitle

\begin{abstract}
The Anderson transition in three dimensions  in a
randomly varying magnetic flux
is investigated in detail by means of
the transfer matrix method with high accuracy.
Both, systems with and without an additional random scalar potential
are considered.
We find a critical exponent of $\nu=1.45\pm0.09$ with random scalar
potential. Without it, $\nu$ is smaller but
increases with the system size and extrapolates
within the error bars to a value close to the above.
The present results support the conventional
classification of universality classes due to symmetry.

\end{abstract}


\narrowtext

Since the pioneering  work of Anderson\cite{Anderson},
the disorder-induced metal-insulator transition, which is one of
the most fundamental quantum phase transitions in condensed
matter physics,
has attracted considerable attention\cite{LR,KM}.
Depending on the symmetry,
the critical behavior of this Anderson transition(AT) is conventionally
classified into three universality classes: the orthogonal,
the unitary and the symplectic class\cite{Dyson}.
Systems invariant under spin rotation as well as 
under time reversal belong to
the orthogonal class.
Unitary systems are
characterized by the absence of time-reversal symmetry, due to, for
instance, a magnetic field. Systems without spin rotation invariance
belong to the symplectic class.

The AT in a homogeneous magnetic field has been studied extensively for
many years, mainly in connection with the quantum Hall effect
\cite{Hajdu}.
In two dimensions(2D) in the presence of a strong magnetic field,
the AT is marginal. States at the centers of the Landau bands are critical
and all the other are localized. At the band centers,
the localization length diverges with the exponent $\nu \sim 2.4$
\cite{Huckestein}.
In 3D, there exist extended states and
the AT takes place \cite{OKO,HKO,CD}. The latter
has recently been re-analyzed and
the critical exponent for the localization
length has been determined to be $1.43 \pm 0.06$\cite{SO}.

When the magnetic field is uniform in space,
randomness is introduced by a  random scalar potential.
On the other hand, in recent years, there has also been considerable
interest in 2D systems subject to a spatially  random  magnetic
field, mainly in connection with  the fractional quantum Hall
effect\cite{RMG}.
The random magnetic field introduces randomness
as well as the absence of invariance under time reversal in a system.

In 3D, the AT in the presence of
a random vector potential and without
a random scalar potential, has been
investigated numerically. The data suggested \cite{OOK} that
the mobility edge is very close to the band edge. The exponent for
the localization length has been estimated to be
$\nu \approx 1$ \cite{OOK} which is
considerably smaller than that in the case with an additional
random scalar
potential and in a uniform magnetic field.
It has also been reported \cite{HKO,KOH} that in the presence of
a random scalar potential, the
critical exponent has a universal value,
irrespective of whether the magnetic field
is uniform or random.
This seemed to indicate that the AT in
a random vector potential but without random scalar potential
is different from the one with
a random scalar potential.
It should be noted that this would question  the validity of
the above conventional classification of the AT because
in both cases the time reversal symmetry is broken and
hence these two systems should belong to the same, namely 
the unitary
universality class.

The critical exponent $\nu \approx 1$ for
the 3D system with a random magnetic field
has been obtained by the finite-size
scaling method \cite{MK}.
This 
method
has been applied successfully to analyze the critical
behavior of the
AT \cite{KM}.
In most cases, however, the numerical analyses
have been restricted to energies  near the band center.
It has been reported that
systematic scaling behavior has not been clearly
observed for energies away from the band center \cite{KBMS,SK}.
For the 3D system with a  random magnetic field, the mobility edge
lies quite close to the effective band edge \cite{OOK}.
It is therefore imperative to investigate the present
problem with considerably higher accuracy and to examine carefully
whether or not the scaling behavior is modified by adding
a random scalar potential.

Recently, high-accuracy scaling analyses of the Anderson transition
have been performed by several authors \cite{SO,MacKinnon}.
It has been concluded \cite{SO} that,
by reducing the errors of raw data to $0.1 \sim 0.2\%$,
 one can numerically distinguish
the unitary from the orthogonal class,
which was impossible when only low accuracy data with $\sim 1\%$
accuracy were used.

Encouraged by this recent success, we have started a numerical
high-precision finite size scaling project, in order to clarify
the above mentioned discrepancy between the critical exponent of the
AT far away from the band center induced solely by randomness
in a vector potential and the exponent obtained for other unitary
systems.

We found a clear systematic dependence of the exponent on the
system-size in the former case which would introduce 
corrections to scaling.
These would become smaller than the statistical error only for system
sizes larger than those which are presently achievable. We estimate
the  asymptotic limit for the exponent to be $\nu \sim 1.4$,
when the transition is near the band edge. This behavior is not 
changed
significantly
when shifting the mobility edge by adding a 
weak
random scalar potential.

The model is defined
by the Hamiltonian \cite{OOK}
\begin{equation}
 H = t \sum_{<i,j>} \exp ({\rm i}\theta_{i,j}) C_i^{\dagger}C_j +
     \sum_i V_i C_i^{\dagger}C_i ,
\end{equation}
where $C_i^{\dagger}(C_i)$ denotes the creation(annihilation)
operator of an
electron at the site $i$ of a 3D cubic lattice.
Energies $\{ V_i\}$ denote
the random scalar potential distributed independently and
uniformly in the range
$[-W/2, W/2]$. The Peierls phase factors
$\exp ({\rm i}\theta_{i,j})$
describe a random vector potential or
magnetic field.
We confine ourselves to
phases $\{\theta_{i,j} \}$ which
are distributed independently and uniformly
in $[ -\pi ,\pi ]$.
The hopping amplitude $t$ is assumed to be the energy unit, $t=1$.

We consider
quasi-1D systems with cross section
$M \times M$ \cite{MK,MacKinnon}.
The Schr\"{o}dinger equation $H \psi = E \psi$
in such a bar-shaped system
can be rewritten using transfer matrices $T_n(2M^2 \times 2M^2)$
\begin{equation}
  \left(   \begin{array}{c}
               \psi_{n+1} \\
               \psi_n
           \end{array} \right)
 = T_n \left(   \begin{array}{c}
               \psi_{n} \\
               \psi_{n-1}
           \end{array} \right) ,\quad
   T_n = \left(   \begin{array}{ll}
               E-H_n  & -I \\
                I    &  0
           \end{array} \right)
\end{equation}
($n=1,2,\ldots$)
where $\psi_n$ and $H_n$ denote the set of coefficients of
the state $\psi$
and the Hamiltonian of the $n-$th slice, respectively.
The identity matrix is denoted by $I$.
The off-diagonal parts of the
transfer matrix $T_n$ can be expressed by the identity matrix
because the phases in the transfer-direction can be
removed by a gauge transformation \cite{OOK}.
The logarithms of the eigenvalues of the limiting matrix $T$
\begin{equation}
 T \equiv \lim_{n \rightarrow \infty} [(\prod_{i=1}^n T_i)^{\dagger}
 (\prod_{i=1}^n T_i)]^{1/2n}
\end{equation}
are called the Lyapunov exponents.
The smallest Lyapunov exponent $\lambda_M$ along the bar is estimated
by a technique which uses the product of these transfer
matrices \cite{KM,MK}.
The relative accuracies for the smallest Lyapunov exponents
achieved here are $0.2\%$ for $M \le 10$ and
$0.25\% \sim 0.3\%$ for $M=12$.
The localization length $\xi_M$ along the bar is given by the inverse of
the smallest Lyapunov exponent, $\xi_M =1/ \lambda_M$.

The assumption of one-parameter scaling for
the renormalized localization length $\Lambda_M \equiv \xi_M /M$
implies
\begin{equation}
 \Lambda_M = f(\xi / M),
\end{equation}
where $\xi=\xi(E,W)$ is the relevant length scale in the limit
$M \rightarrow \infty$\cite{MK}.
Near the mobility edge $E_c(W)$, $\xi$ diverges with an exponent $\nu$
as $\xi \sim x^{-\nu}$ with
$x=(E-E_c)/E_c$. If the transition is driven by the disorder
$W$ at a constant energy,
$x=(W_c-W)/W_c$.
At the mobility edge, $\Lambda_M$ becomes
scale-invariant. The quantity $\Lambda_M$ is a smooth
function of $E$ and $W$, and we can expand it
as a function of $x$ as
\begin{eqnarray}
 \Lambda_M &=& \Lambda_c +
 \sum_{n=1}^{\infty} A_n (M^{1/\nu}x)^n .
 \label{fitcur}
\end{eqnarray}
By fitting our data to the above function,
we can determine the critical exponent $\nu$ and the mobility edge
accurately.
In practice, we truncated the series (\ref{fitcur}) at the third order.

We used the standard $\chi^2$-fitting procedure\cite{For}.
In order to check the goodness of the fit,
we also evaluated the probability $Q$ that the $\chi^2$ will
exceed the minimum value $\chi_{\rm min}^2$
actually obtained by the
fit. The probability $Q$ is evaluated via the
incomplete gamma  functions and
the condition $Q > 0.001$ is often regarded
as an acceptable condition for the fitting function \cite{For}.
If the value of $Q$ is too small,
in other words, the minimum
value of the $\chi^2$ is considerably large,
there may be
systematic deviations of the numerical data from the fitting function.
In the recent work \cite{SO}
on the Anderson transition in 3D orthogonal
and unitary systems,
it has been demonstrated that the above fitting function
up to the third order
is in fact valid.
The error bars are estimated by using the Hessian matrix
and the confidence interval is chosen to be
$95.4\%$.

We consider first the AT at
the band center $E=0$ in the presence of a strong
random scalar potential as well as random vector potential.
The renormalized localization lengths
$\Lambda_M$ evaluated for the disorder $W$ in the range
$ [17.8, 19.8]$ and sizes $M=6,8,10$ and $12$ are shown in
figure 1.
The above described fit yields
the critical exponent $\nu = 1.45 \pm 0.09$ and
the critical disorder $W_c = 18.80 \pm 0.04$.
The renormalized localization length $\Lambda_c$ at the critical point
is $0.558 \pm 0.003$.
The value of $Q$ is
$\sim 0.89$ which confirms the validity of the fitting function
(\ref{fitcur})
and thus of the one-parameter scaling behavior of $\Lambda_M$ in this
range of the disorder $W$.
The error bars of the present results are at least a factor
of 3 smaller than those of the previous
estimates \cite{HKO}.
\begin{figure}[t]
\epsfxsize=3.1in \epsfbox{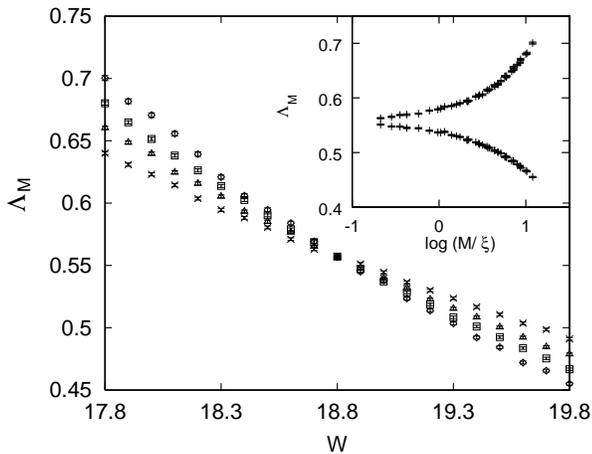}
\caption{The renormalized localization length $\Lambda_M$ as a
function of disorder $W$ for different sizes.
The crosses, triangles, the squares
and the diamonds correspond to $M=6$, $8$,
$10$ and $12$, respectively. Inset: The scaling function.}
\end{figure}

Next, we concentrate on the AT in  the random magnetic
field {\it but} without random scalar potential $(W=0)$.
In the previous work \cite{OOK}, the energy range used for
the scaling analysis was assumed to be $4.3 \le |E| \le 4.5$. This
is very close to the band edge and thus the density of
states(DOS) is rapidly decreasing \cite{OOK}.
To get rid of the influence by this rapid change of the
DOS, the energy window
for the scaling analysis should be taken to be as small as possible.
We therefore choose calculated data for
$4.39 \le E \le 4.44$. The energy window is 4 times smaller than that in
\cite{OOK}.
The numerical data for $\Lambda_M$ are shown in figure 2.
The transition can be located around $E \approx 4.415$.
We then fit the data for $M=6,8,10$ and $12$ to the function
(\ref{fitcur}). This yields
$E_c \approx 4.414$,  $\nu \approx 1.2$ and
$\Lambda_c \approx 0.51$, which is consistent with the
previous estimates \cite{OOK}. It is found, however, that
in the present case, the value of $Q$ turns out to be very small
, namely $\sim 10^{-14}$,
in contrast to the above considered case of the band center.
This shows that systematic deviations of the data from
the fitting function are very likely to exist.
We therefore have to analyze the  numerical data much more carefully.

In order to get insight into the origin of the deviations of
the data from
the fitting function,
we carried out the fits using different combinations of
system-sizes $M_1$ and $M_2=M_1 +2$ for
$M_1 =6,8$ and $10$ (table I).
Although the crossing point is
almost size independent, the exponent shows a systematic dependence
on the system cross section.
For $M=6$ and $8$, in particular, the exponent is close to 1,
while for the other two cases it is around 1.3 which is
close to the value estimated for a system in a uniform magnetic field.
In addition, for $M=6$ and $8$ the value of $Q$ becomes much smaller
than those for the other two cases, which could imply that the fitting
function
is not working well for these smaller sizes.
\begin{figure}
\epsfxsize=3.1in \epsfbox{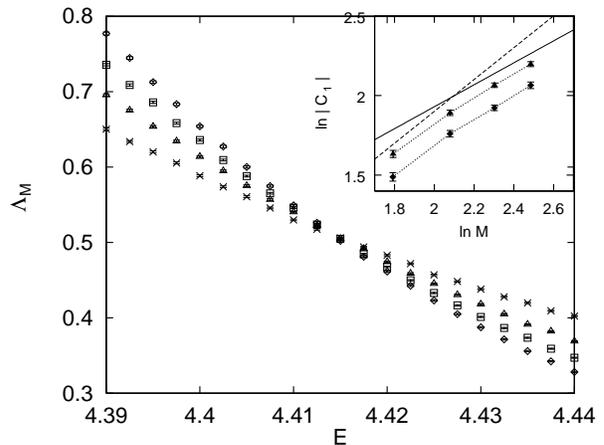}
\caption{The renormalized localization length $\Lambda_M$ as a
function of energy $E$ for $W=0$. The crosses,
triangles, the squares
and the diamonds correspond to $M=6$, $8$,
$10$ and $12$, respectively.
Inset: The logarithm of the derivative $C_1 \equiv d\Lambda_M/dE$
for $W=0$ at
$E=4.414 \approx E_c$(filled triangles)  and for $W=1$ at
$E=4.451 \approx E_c$(filled diamonds)
as a function of $\ln M$. The solid line and the
dashed lines represent the slope with $\nu=1.45$ and $\nu=1$,
respectively.}
\end{figure}

We also analyzed the data in
a different way.
We performed a third order polynomial fit for each size and
estimated the derivative $C_1 \equiv d\Lambda_M /dE |_{E=E_c}$
at the critical energy
$E_c=4.414$. The value of $Q$ in this case is larger than
$0.1$ for any size, which means that the 3rd order polynomial fit
itself is working fairly
well. We  plot  $\ln |C_1|$ as a function
of $\ln M$ (figure 2, inset). If scaling works,
the slope is related to the exponent by $1/\nu$.
It is clear from figure 2 (inset)
that the slopes deviate from 1 and are likely to
approach the value $\nu \sim 1.45$, when $M$ is increased.
These two analyses suggest that the present system-sizes may not
be large enough to observe clear one-parameter scaling behavior
for $W=0$.

To see whether these features are specific to $W=0$,
we performed the same analysis for a system
with weak scalar randomness $W=1$. The results of the
fits using different combinations of system-sizes and
by the third order polynomial fits
are listed in table II and shown in figure 2 inset, respectively.
In table II, we see that the exponent for $M=6$ and $8$
again turns out to be close to $1$
and deviates from those for the other two cases,
although the crossing points are rather stable with respect to the
change of sizes. It should also be noted that the value of $\Lambda_c$ is
almost the same as for $W=0$.
In the inset of figure 2, we can observe again the deviation
of the slopes from 1 as the size is increased.
Thus, the system
with a weak  random scalar potential shows
similar behavior as that without a random scalar
potential.

The results for $W=1$ and $W=0$ indicate that the 
correction to scaling is not specific to the case $W=0$. It is 
natural to expect that this is
due to 
the fact that 
the transition is near the band edge.
In both cases, $W=1$ and $W=0$,
the fits using the size $M=6$ give smaller values
of the critical exponent $\nu$. We conclude that this is
the reason why the exponent found in the
previous work \cite{OOK} was smaller.
The present analysis for $W=1$ and $W=0$ also shows
that finite-size corrections to  scaling
exist for the presently achievable system sizes ($6 \le M \le 12$),
which might be the reason of the discrepancy between $\Lambda_c$ near
the band edge and that at the band center.

In summary, we have re-investigated in detail the Anderson transition
in the random magnetic field with and without
random scalar potential. We have evaluated the localization length
along quasi-1D systems
with high accuracy and examined
the scaling behavior of
the renormalized localization length near the transition.
We have confirmed the one-parameter scaling behavior for the transition
at the band center with a relatively strong random scalar
potential  and found
the exponent $1.45\pm 0.09$. This value agrees well with
the recent precise results
for systems in a uniform magnetic field\cite{SO}.
We have also performed the finite-size scaling analysis for both, the
system without random scalar potential $(W=0)$ and
with a weak random scalar potential (W=1).
In both cases, deviations of numerical data
from the scaling ansatz are found, especially for smaller sizes.
As the size is increased,
the exponent is
more likely to approach to a value around $1.4$
rather than the value of 1.
In particular, no evidence for the exponent $1$ has been found.
 
On the basis of the present results, we conclude
that there exists no evidence that the critical behavior
in a 3D system in a random magnetic field
is different from that for other unitary systems. 
This supports  
the conventional classification of the AT by
universality classes due to symmetry.

The authors thank M. Batsch, A. MacKinnon and I. Zharekeshev
for valuable discussions.
The numerical calculations were done on a FACOM VPP500 of
Institute for Solid State Physics, University of Tokyo and
in computer facilities of I. Institut f\"{u}r Theoretische Physik,
Universit\"{a}t Hamburg.
This work was supported in part by the EU-project FHRX-CT96-0042
and by the Deutsche Forschungsgemeinschaft
via Project Kr627/10 and the Graduiertenkolleg
``Nanostrukturierte Festk\"{o}rper''.
One of the authors (T.K.) thanks Alexander von Humboldt Foundation
for financial support during his stay at University of Hamburg.


%
\begin{table}
\begin{tabular}{ccccc}
 $(M_1,M_2)$ & $\Lambda_c$ & $\nu$ & $E_c$ & $Q$ \\ \hline
$(6 , 8)$ & $0.514\pm 0.005$ & $1.05\pm 0.07$ & $4.414\pm 0.001$ &
$\sim 10^{-5}$ \\
$(8 , 10)$ & $0.516\pm 0.007 $ &$1.26\pm 0.09 $ &$4.414\pm 0.001 $ &
$\sim 0.89$ \\
$(10 , 12)$ &$0.51\pm 0.01 $ & $1.32\pm 0.12 $ &$4.414\pm 0.001 $ &
$\sim 0.99 $
\end{tabular}
\caption{Results of the fits for different sizes in the absence of
random scalar potential $(W=0)$.}
\end{table}
\begin{table}
\begin{tabular}{ccccc}
$(M_1,M_2)$ & $\Lambda_c$ & $\nu$ & $E_c$ & $Q$ \\ \hline
$(6 , 8)$ & $0.510\pm 0.005$ & $1.09\pm 0.08$ & $4.451\pm 0.001$ &
$\sim 0.91$ \\
$(8 , 10)$ & $0.519\pm 0.008 $ &$1.36\pm 0.12 $ &$4.450\pm 0.001 $ &
$\sim 0.56$ \\
$(10 , 12)$ &$0.51\pm 0.01 $ & $1.34\pm 0.14 $ &$4.452\pm 0.002 $ &
$\sim 0.95 $
\end{tabular}
\caption{Results of the fits for different sizes in the presence of
the weak random scalar potential $(W=1)$. The energy window
is taken to be $4.425 \le E \le 4.475$ and the same number of
data points as the case of $W=0$ are used for the scaling analysis.}
\end{table}

\begin{references}
\bibitem{Anderson} P.W. Anderson, Phys. Rev. {\bf 109}, 1492 (1958).
\bibitem{LR} P.A. Lee and T.V. Ramakrishnan, Rev. Mod. Phys. {\bf 57},
             287 (1985).
\bibitem{KM} B. Kramer and A. MacKinnon, Rep. Prog. Phys. {\bf 56},
             1469 (1993).
\bibitem{Dyson} F.J. Dyson, J. Math. Phys. {\bf 3}, 140 (1962);
                {\bf 3}, 157 (1962); {\bf 3}, 166 (1962).
\bibitem{Hajdu} {\it Introduction to the Theory of the Integer
Quantum Hall Effect}, J. Hajdu {\it et al.},
             (VCH, 1994).
\bibitem{Huckestein} B. Huckestein, Rev. Mod. Phys. {\bf 67}, 357 (1994).
\bibitem{OKO} T. Ohtsuki, B. Kramer and Y. Ono, J. Phys. Soc. Jpn.
              {\bf 62}, 224 (1993).
\bibitem{HKO} M. Henneke, B. Kramer and T. Ohtsuki, Euro. Phys. Lett.
              {\bf 27}, 389 (1994).
\bibitem{CD} J.T. Chalker and A. Dohmen, Phys. Rev. Lett. {\bf 75},
 4496 (1995).
\bibitem{SO} K. Slevin and T. Ohtsuki, Phys. Rev. Lett. {\bf 78}, 4083
             (1997).
\bibitem{RMG} B.I. Halperin, P.A. Lee, and N. Read, Phys. Rev.
              {\bf B47}, 7312 (1993).
\bibitem{OOK} T. Ohtsuki, Y.Ono and B. Kramer, J. Phys. Soc. Jpn.
             {\bf 63}, 685 (1994).
\bibitem{KOH} B. Kramer, T. Ohtsuki and M. Henneke, in {\it
              Quantum Dynamics of Submicron Structures}, edited by
              H.A. Cerdeira et al., 21
              (Kluwer Academic Publishers, 1995).
\bibitem{MK} A. MacKinnon and B. Kramer, Phys. Rev. Lett. {\bf 47}, 1546
             (1981); Z. Phys. B {\bf 53}, 1 (1983).
\bibitem{KBMS} B. Kramer, K. Broderix, A. MacKinnon and M. Schreiber,
              Physica A {\bf 167}, 163 (1990).
\bibitem{SK} M. Schreiber and B. Kramer, 
             in {\it Anderson Transition} edited
             by T.Ando and H.Fukuyama, 92 (Springer-Verlag 1988).
\bibitem{MacKinnon} A. MacKinnon, J. Phys.: Condens. Matter {\bf 6}, 2511
              (1994).
\bibitem{For} W.H. Press {\it et al.}, 
              Numerical Recipes (Cambridge University
              Press, 1986).
\end{references}
\end{document}